\documentclass[12pt]{iopart}
\usepackage{graphicx}
\usepackage{wrapfig}
\usepackage{color}

\begin{document}
\title[Bose-Einstein condensation in large time-averaged optical ring potentials]{Bose-Einstein condensation in large time-averaged optical ring potentials}
\author{Thomas A Bell$^{1,2}$, Jake A P Glidden$^{1,2}$, Leif Humbert$^1$, Michael W J Bromley$^1$, Simon A Haine$^{1}$, Matthew J Davis$^{1,3}$, Tyler W Neely$^{1,2}$, Mark A Baker$^{1}$, Halina Rubinsztein-Dunlop$^{1,2}$}

\address{$^1$ School of Mathematics and Physics, University of Queensland, St Lucia, QLD 4072, Australia}
\address{$^2$ ARC Centre of Excellence for Engineered Quantum Systems (EQuS), School of Mathematics and Physics, University of Queensland, St Lucia, QLD 4072, Australia}
\address{$^3$ JILA, University of Colorado, 440 UCB, Boulder, CO 80309, USA}
\ead{m.baker@physics.uq.edu.au}

\begin{abstract}

Interferometric measurements with matter waves are established techniques for
sensitive gravimetry, rotation sensing, and measurement of surface interactions, but
compact interferometers will require techniques based on trapped geometries. In a step
towards the realization of matter wave interferometers in toroidal geometries, we
produce a large, smooth ring trap for Bose-Einstein condensates using rapidly scanned 
time-averaged dipole potentials. The trap potential is smoothed by using the atom distribution as input to an optical intensity correction algorithm. Smooth rings with a diameter up to 300 $\mu$m are demonstrated. We experimentally observe and simulate the
dispersion of condensed atoms in the resulting potential, with good agreement serving as
an indication of trap smoothness. Under time of flight expansion we observe low energy
excitations in the ring, which serves to constrain the lower frequency limit of
the scanned potential technique. The resulting ring potential will have applications as a
waveguide for atom interferometry and studies of superfluidity.
\end{abstract}

\submitto{\NJP}
\maketitle

\section{Introduction}

The development of tailored geometries and novel potentials for dilute gas Bose-Einstein condensates (BECs)\cite{Ande95,Davi95,Brad95} has both expanded  our fundamental understanding, as well as providing a path to exploiting their unique properties, such as coherence, for the enhancement of precision measurement.  One of the geometries of interest is the ring trap, which can serve as a waveguide with an enclosed area for atom interferometry \cite{Shin04,Schumm05,Garc06}, for studies of superfluidity \cite{Rye07,Beat13,Jend14,Eckel14,Eckel14a}, a storage device for atomtronic circuits \cite{Seam07}, continuously connected ring-shaped optical lattices \cite{Amic05,Amic14}, and development of a BEC analog of SQUID circuits \cite{Rama11,Rye13,Math15}. In the past few years there has been considerable activity in this area, and a number of techniques have been developed and demonstrated.

Magnetic ring traps \cite{Saue01,Gupt05,Arno06}, including those utilising RF dressed states \cite{Mori06,Lesa07,Sher11}, have the advantage that the current generating elements can be very far away from the atomic system, resulting in intrinsically smooth potentials. However, their utility is limited to holding and guiding magnetically trappable states, with lifetimes subject to Landau-Zener losses, and the potential landscapes that can be generated are limited in scope and difficult to change dynamically.

More recently, traps based on optical forces have been demonstrated, making use of the attractive (repulsive) dipole interaction with red (blue) detuned light to shape ring shaped potentials. A variety of techniques have been developed, including Laguerre Gaussian beams \cite{Rama11,Beat13,Wright00}, time-averaged optical potentials \cite{Schnelle08,Hend09,Roberts14}, conical refraction from optically biaxial crystals \cite{Turp15}, overlapped blue and red detuned beams \cite{Mart15}, and more recently employing digital micromirror devices \cite{Kuma15}. Optical traps have the advantage of being state insensitive, allowing the trapping and manipulation of spinor condensates \cite{Stam98,Barr01,Stam13}.

Thus far, the majority of optical traps have been limited to small ring diameters less than 50 $\mu$m,  where the larger mean field of the condensate in the smaller volume is able to overcome any azimuthal roughness of the potential.  Confinement in the vertical direction is usually supplied by a far detuned Gaussian TEM$_{00}$ (red detuned) or Hermite Gaussian TEM$_{01}$ (blue detuned) light sheet. However, the potential depth changes across the sheet, and is defined by the Gaussian profile in the transverse direction, and the Rayleigh length in the longitudinal direction. The result is a saddle-like variation of the trap potential at large ring diameters.   For applications such as matter wave interferometry, where the phase shift is a function of enclosed area, larger ring diameters are desirable. 

We present here details of our BEC apparatus, where we use a rapidly scanned red detuned dipole beam to produce smooth homogenous dipole potentials. The beam is scanned using a 2D acousto-optical deflector (2D-AOD) and allows us to generate versatile trapping geometries, and in the context of this work, ring traps for BEC. We show that using an intensity correction system, which uses the atom density as a probe of imperfections, we can correct for optical aberrations and produce large homogeneous ring traps suitable for studies of superfluidity and atom interferometry using BEC. We study the dynamics and  lifetime of the BEC and ring, and demonstrate the application of our ring trap as a smooth circular waveguide for atom interferometry. Finally, we study the expansion of the ring under time of flight, and observe phonon excitations. 

\section{BEC production}

Our experimental BEC apparatus uses $^{87}$Rb, and consists of a two vacuum cell arrangement of Rb source and science cell.  In the first rectangular glass cell, maintained with a Rb vapour pressure of $10^{-7}$ mbar, a 2D-MOT is generated which forms the beam source of cold atoms for the secondary UHV octagon shaped glass science cell, located 350 mm downstream.  The two individually pumped cells are separated by a machined copper rod with a 1.5 mm diameter aperture of 12 mm depth, maintaining a factor $\approx$ 1000 pressure differential. In the science cell, we form a 3D MOT with $2\times{10^{9}}$  atoms in 10 s. The MOT light is then extinguished and the atoms are loaded and compressed in a quadrupole magnetic trap, with $N=1 \times{10^{9}}$ atoms at temperature of 120 $\mu$K, and axial gradient of ${dB}/{dz} = 160$ G cm$^{-1}$. RF and optical evaporation is then performed in a hybrid magnetic trap and single beam optical dipole trap ($\lambda = 1064$ nm, $w_0 = 65$ $\mu$m the $1/e^2$ beam waist), closely following the sequence outlined in \cite{Lin09}. This robust technique combines the advantages of the respective trapping techniques, with large initial capture volume and simplicity of RF evaporation provided by the quadrupole magnetic trap, and fast final evaporation stage in the dipole trap. With this, we produce Bose-Einstein condensates of $^{87}$Rb  in the $F=1$, $m_F=-1$ state, with pure condensates up to $N=1 \times{10^{6}}$ atoms.  Our final trap parameters are typically $P=100$ mW, an axial gradient field of ${dB}/{dz}=28$ G cm$^{-1}$ giving trap frequencies of $(\omega_x,\omega_y, \omega_z) = 2 \pi \times (78,72,29)$ Hz.\\
\\
Imaging of our BEC is performed in the vertical plane using a four element compound objective \cite{Benn13}, with f=47 mm, and corrected for the 4 mm quartz windows on the science cell. The objective provides an optical resolution of 1.6 $\mu$m, calibrated from the point spread function (PSF) of a pinhole aperture. A secondary AR coated achromat lens is used to form the magnified image onto the CCD camera (ProSilica EC1380), resulting in a magnification of M = 6.38 (f = 300 mm), or M = 15.95 (f = 750 mm).

\section{Time-averaged optical potentials}

\begin{figure}[h]
\begin{center}
\includegraphics[width=0.5\textwidth]{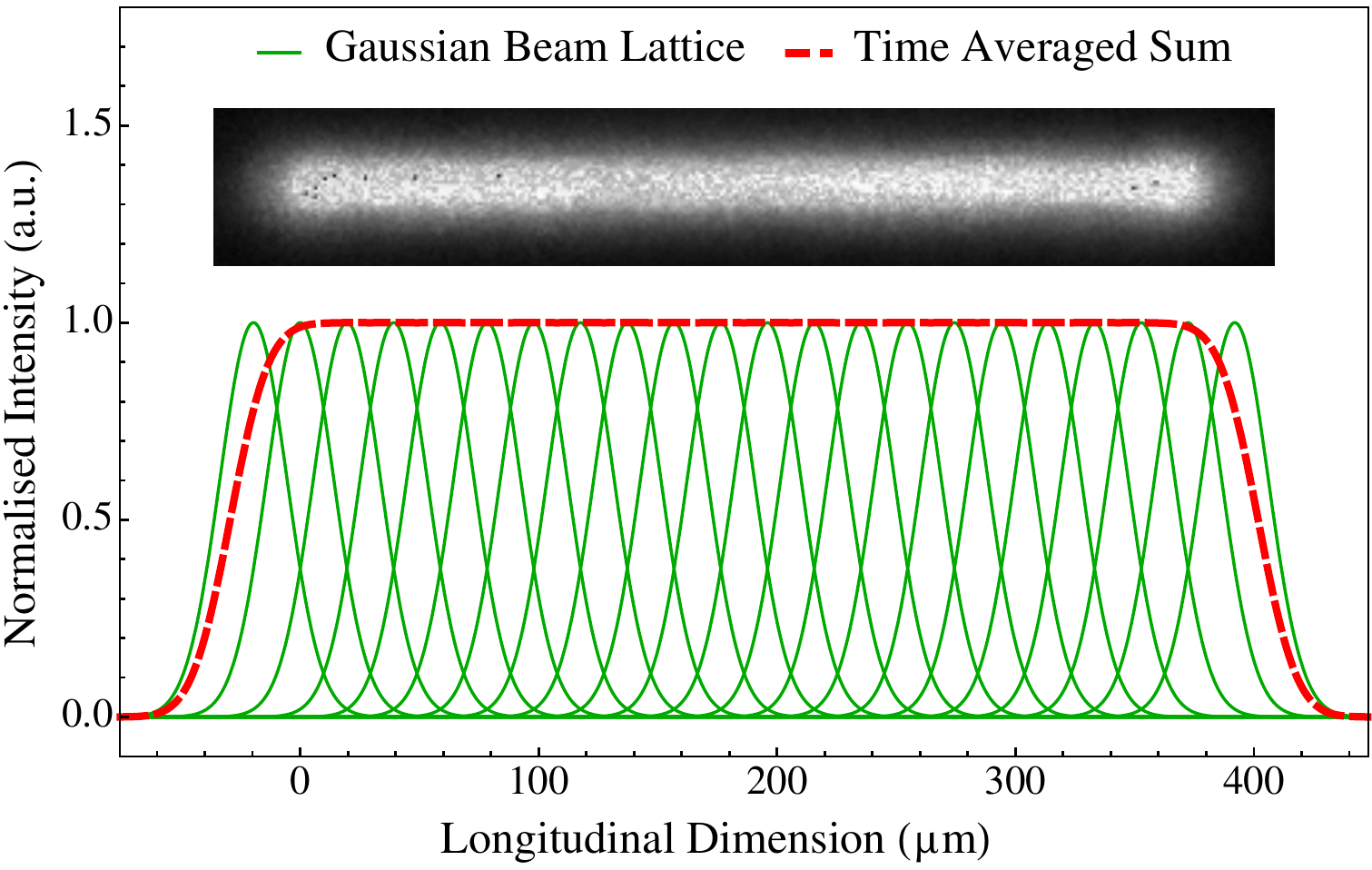}
\end{center}
\caption{ Time-averaged potential, formed by rapidly scanning between each of the positions of the Gaussian beam, spaced by 0.7$w_s$. In this example, 22 points are overlapped in a line, resulting in a line-trap of 370 $\mu$m length. (Inset) Resulting in-trap absorption image of the $^{87}$Rb condensate in the time-averaged line trap, for a scan rate of 20 kHz. \label{fig1}}
\end{figure}

The scan potentials are made of the time-averaged sum of discrete overlapping Gaussians, controlled using the 2D-AOD, where neighbouring beam positions are spaced by 0.7$w_s$, where $w_s$ is the $1/e^2$ radius of the scanning beam (Fig.~\ref{fig1}). This spacing results in a corrugation free potential, and maximises scan bandwidth by minimising points, and allows us to have individually addressable points to correct for intensity aberations. In our present optical configuration, with a 200 mm focusing lens, $w_{s} = 28$ $\mu$m. The number of points required for a given length or arc $L$ is $N_s={L}/(0.7 {w_s})$. The scanning bandwidth of the 2D-AOD is defined by the access time, which is the transit time of the acoustic wave across the beam waist of a beam traversing the AO crystal. It is defined as $\tau_a={d}/{v_s}$ where $d$ is the beam diameter and $v_s$ is the speed of sound in the crystal. For our AOD, which is optimised for the shear mode of a TeO$_2$ crystal with $v_s = 617$ ms$^{-1}$, $\tau_a=1.6$ $\mu$s for a 1 mm beam. The maximum scan rate for a pattern with $N_s$ points  is then $f_s={1}/({N_s \tau_a})$. For our system, this sets a practical limit of around 20 kHz.  For scan rates that exceed this criteria, clean switching between resolvable points is compromised, resulting in smearing or blurring of the pattern. A higher bandwidth can be achieved by reducing the beam diameter in the crystal, or using longitudinal modes in the crystal with higher speed of sound. \\

\section{Scanning potential setup}

\begin{figure}[h]
\begin{center}
\includegraphics[width=0.75\textwidth]{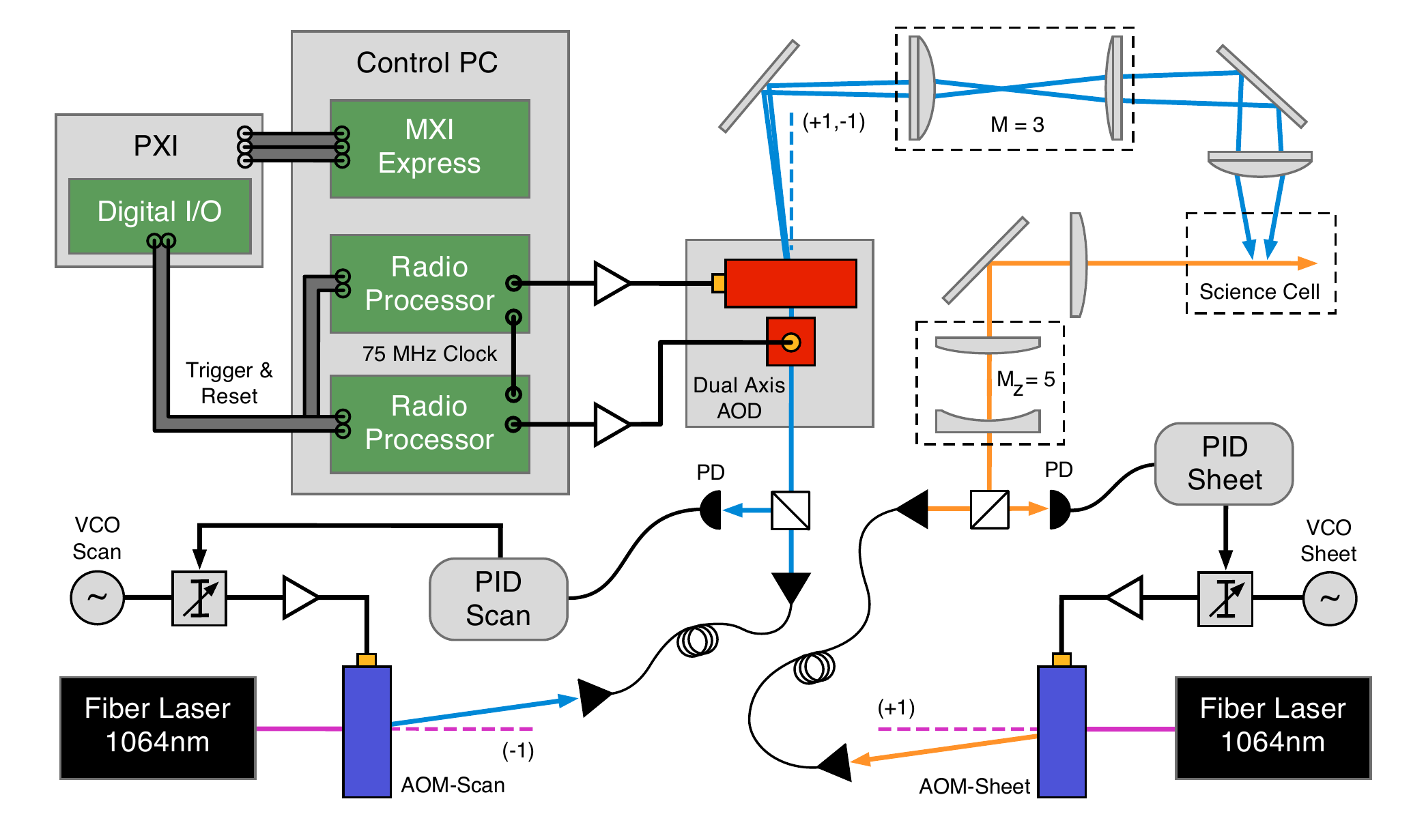}
\end{center}
\caption{Optical setup for time-averaged potentials. The scan beam and sheet potential intensity are independently controlled using AOM-scan and AOM-sheet, and intensity locked using a photodiode (PD) and PID feedback circuit. The scan image is generated using the (1,-1) order of the 2D-AOD, and the image is magnified ($\times$3) before focusing into the science cell.  \label{fig2}}
\end{figure}

The setup for our scanning trap potentials is shown in Figure 2. The scanning potential is generated from a 1064 nm multimode fibre laser (IPG Photonics, YLR-5-1064-LP). The first diffracted order from a single fixed frequency acousto-optical modulator (AOM-scan, $f=80$ MHz) is coupled to a single mode polarization maintaining fibre. A photodiode placed after the fibre is used as the input to a feedback PID circuit and referenced to a control voltage. The PID output drives a voltage variable attenuator which sets the AOM RF power and provides intensity stabilisation and control. 

The fibre output has a diameter of 2 mm, and is then passed through the 2D acousto optical deflector (2D-AOD) (IntraAction DTD-274HA6). The deflection angle of the beam in the AOD is controlled by the frequency of the traveling radio frequency RF wave inside the crystal. Modulating the RF frequencies allows the beam to be scanned in arbitrary 2D patterns. The AOD operates with central RF frequency of 27 MHz, has an active aperture of 4 mm x 4 mm, and is driven from two separate computer controlled and amplified RF cards (Spincore RadioProcessor), with one card per axis. The RF cards are synchronized to a common 75 MHz clock, and capable of better than 100 ns resolution. The 2D-AOD is mounted on a tilt optical stage, and the alignment is optimised for the $(1,-1)$ order, with typically more than 55$\%$ efficiency. The output beam from this order is then further expanded through a $3\times$ magnification telescope, and then imaged vertically into the science cell with a 200 mm achromatic lens. The resulting beam has a waist of $w_s=28$ $\mu$m, and the position and amplitude of the beam is controlled from the RF frequency and amplitude registers of the RF cards driving the 2D AOD. By rapidly scanning through these registers at a rate faster than the trapping frequencies, the atoms experience the time-averaged potential.

Confinement of the atoms against gravity in the vertical plane is provided by an overlapped light sheet potential, also shown in Figure 2. The light sheet is controlled from a single pass AOM (AOM-sheet)and intensity controlled with a PID circuit.   The fibre output is first cylindrically expanded in a $5\times$ magnification telescope, and then focused into the science cell with a final f = 300 mm cylindrical lens. The sheet potential has a measured horizontal waist $w_{x,\mathrm{sheet}}=1.25$ mm, $w_z=27$ $\mu$m, and Rayleigh length $y_{R} = 2.1$ mm. For an optical power of 600 mW, this results in a vertical trap frequency of $\omega_{z} = 2{\pi} \times 110$ Hz. The scan beam and sheet beam are set with orthogonal polarization directions to minimise interference effects, and have frequency difference of 160 MHz, set by the respective frequency shift of the intensity control AOMs.

To load the ring trap we begin with the $^{87}$Rb BEC in the hybrid optical and magnetic quadrupole trap, before lowering the dipole trap power while simultaneously ramping up the sheet potential. The atoms are held in the sheet just below the magnetic field minimum.  The ring scan is triggered and ramped in intensity, and the magnetic field gradient is slowly lowered to allow the atoms to fill the ring. In the work presented here, the ring is loaded by the overlap of one edge of the ring with the initial BEC position.

\section{Atom density correction for generating smooth ring-trap potentials}

Using scanning optical potentials, some form of intensity correction is generally a necessary requirement to generate smooth homogenous traps for atoms. The principal factor to be addressed is the variable efficiency of the 2DAOD as the drive frequency is changed, resulting in intensity variation at different deflection angles of the beam. The efficiency response is deterministic, however, as the diffraction efficiency of the AOD is linear with RF drive power, allowing a feedforward scheme where the intensity of the beam is measured and the RF amplitude corrected. We have previously implemented this scheme \cite{Schnelle08} and found it suitable for generating smooth potentials across small areas. For larger patterns of more than 50 $\mu$m, we find the technique to be inadequate, as it does not correct for other issues, such as the Gaussian profile across the sheet potential, or spherical aberrations for off-axis points across the scan region.

Instead, we complement this scheme by using the distribution of the ultra-cold atoms in the trap, measured from absorption imaging, as a sensitive probe for imperfections. By using the atoms as the probe, we measure and apply an intensity correction to the potential, accounting for these other factors. In this way, the potential offsets across the potential can be corrected, albiet at the cost of a modulated radial confinement across the potential.  Although the demonstration of the scheme is performed here for a ring potential, it applies generally to any potential.

We start by initialising the ring potential to have the same amplitude registers at every point. The experimental sequence is run, and the condensate is loaded into the potential. We hold the atoms in the ring for a number of seconds to have confidence that residual excitation of any breathing or sloshing modes from transferring from the hybrid trap to the sheet and scanned potential are minimised. The trapping light is extinguished, and an absorption image of the {\emph{in situ}} density distribution is taken. The absorption image is analysed by dividing the ring trap into segments coinciding with the scanned beam locations, and integrating the atomic density in these areas (Figure 3).  The measured atom number in each of these segments is used as the metric for applying correction to the beam intensity. The algorithm then generates a corrected set of amplitude registers that are fed to the RF cards. The sequence is repeated until the uniformity of the atom density around the ring converges. Typically, after only 5 iterations, the integrated density variations are less than 10 \%. 
Figure 3 shows a typical absorption image of the BEC before and after the correction scheme is implemented. Without correction, the atoms are localised at the initial position, but after correction, the atoms are evenly distributed around the ring.

\begin{figure}[h]
\begin{center}
\includegraphics[width=1\textwidth]{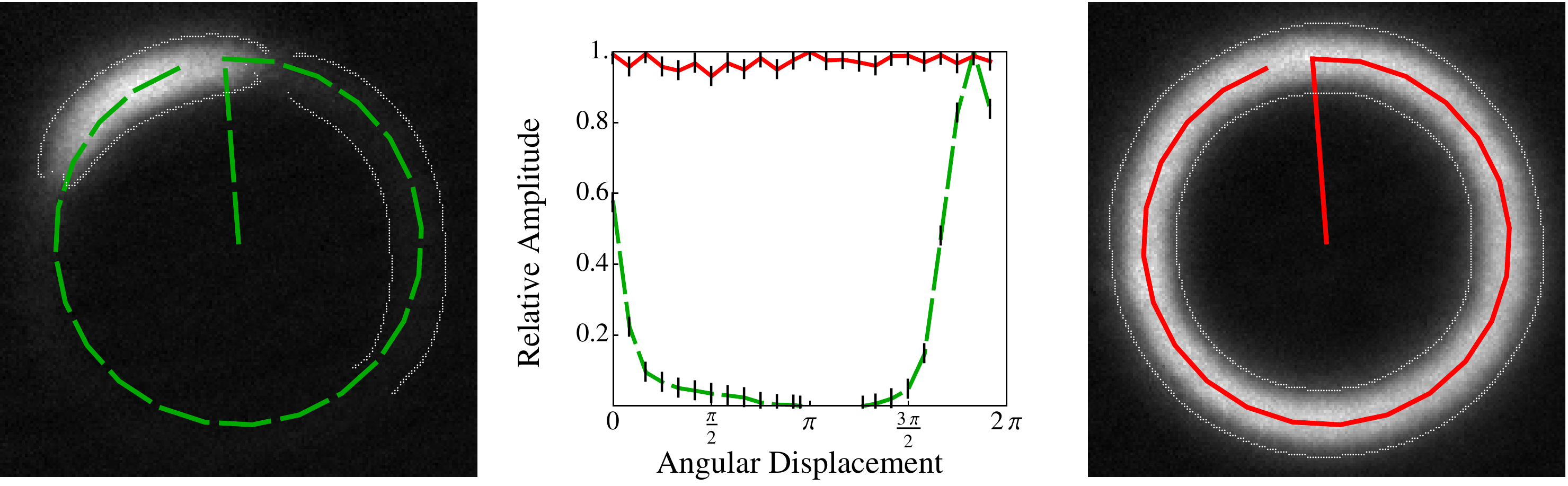}
\end{center}
\caption{{ In-trap absorption images of the BEC, in a 140 $\mu$m diameter ring trap. The white lines are a mathematical fit to determine the ring geometry of the absorption image. The green (red) overlaid line segments on the images indicate the starting and scan position of the beam around the ring. Left: before intensity correction. Right: after intensity correction algorithm has been applied. Centre: Azimuthal integrated density profile around the ring before (green) and after (red) correction.
\label{fig3}}}
\end{figure}

\section{Condensate parameters in a 3D ring trap}

For a 3D ring trap, the potential can be treated as harmonic in the radial and vertical directions, and continuously connected in the azimuthal direction.  The expression for critical temperature $T_c$ takes on a modified form from the familiar 3D harmonic potential \cite{Mori06,Bagn87,Nuge09}

\begin{equation}
T_c={\left({\frac{\sqrt{2}N{\hbar}^3 \omega_{\rho} \omega_z}{1.514 {k_B}^{5/2} m^{1/2} \pi R}}\right)}^{2/5}
\end{equation}

\noindent where $N$ is the number of atoms in a ring of radius $R$, $m$ is the atomic mass, and $\omega_{\rho}$, $\omega_z$ are the radial and vertical trap frequencies. For a given trap geometry, the ground state occupation $N_0$, or condensate fraction $N_0/N$ at temperature $T$ is given by 
\begin{equation}
\frac{N_0}{N}=1-\left(\frac{T}{T_c}\right)^{\alpha}
\end{equation}
 with the power $\alpha$ describing the dimensionality of the potential. For a 3D harmonic potential $\alpha = 3$  and for a 3D ring potential $\alpha =5/2$. For large atom number, where mean-field interactions dominate, and the kinetic energy terms can be neglected (Thomas-Fermi regime), the chemical potential \cite{Mori06} is 
\begin{equation}
\mu=\hbar\sqrt{\frac{2 {N_0} \omega_{\rho} \omega_z{a_s}}{\pi R}}
\end{equation}
\noindent where $a_s$ is the s-wave scattering length. From this expression, we calculate the peak speed of sound in the trap $c_s = \sqrt{\mu/m}$.

\section{Scan rate and BEC lifetime}

\begin{figure}[h]
\begin{center}
\includegraphics[width=0.8\textwidth]{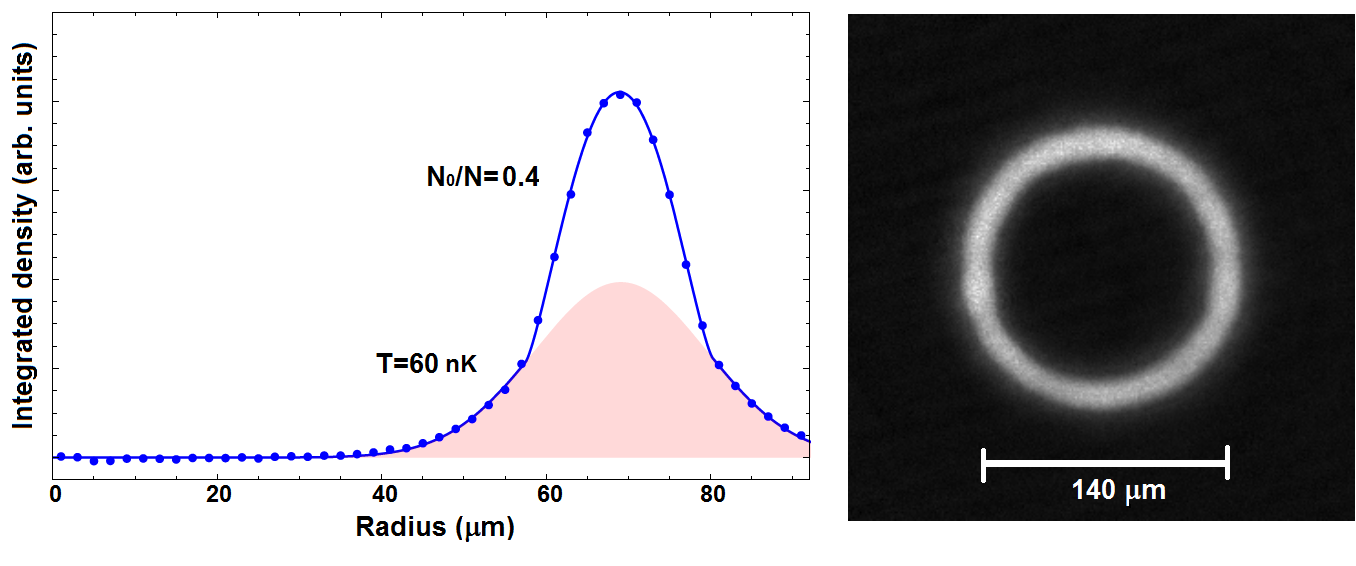}
\end{center}
\caption{(a) Integrated azimuthal profile of the 140 $\mu$m diameter ring trap, showing clear bimodal signature of BEC. The total atom number $N=3.8 \times 10^5$ atoms, $T=60$ nK, with condensate fraction of 0.40. (b)  In-trap absorption image of the ring BEC.\label{fig4}}
\end{figure}

Using a ring diameter of 140 $\mu$m, and ($\omega_{\rho}$, $\omega_{z})=2\pi \times (47,110)$ Hz, we loaded condensates in the compensated ring using a scan rate of 8 kHz. Figure~\ref{fig4} shows $N=3.8\times 10^{5}$ atoms in the ring potential with trapping frequencies $(\omega_{\rho},\omega_z)=2\pi\times(47,110)$ and trap depth of 190 nK.  Using a bi-modal fit to the cloud, we determine the temperature from the thermal fraction to be 60 nK, and measure the condensate fraction to be $N_0/N$=0.40. The calculated critical temperature for the ring is $T_c =75$ nK, giving a theoretical condensation fraction of 0.43 which is in good agreement with that observed. The calculated chemical potential is 10 nK.\\
\\
In order to study the effect of scan rate on lifetime, we varied the scan rate of our ring from 200 Hz to 20 kHz, a ratio of scan rate to trap frequency ${\omega_{scan}}/{\omega_{\rho}}$ $\approx$ 4 -- 400, and held the BEC in the trap for up to 30 s.  We found scan rates of 200 Hz were sufficient to generate continuous ring BEC, but with a lifetime of only 10 s, limited in part by proximity to parametric resonances. For scan rates of 600 Hz or larger, we observed no strong dependence on scan frequency for the lifetime of atom number or condensates, with 1/e lifetimes in excess of 20 s. This is consistent with the typical measured lifetime of the BEC in our hybrid trap, of 25 seconds. Using a higher power (1.5 W) in the sheet potential with $\omega_{z} =2\pi\times 170$ Hz saw a reduction in the lifetime to approximately 6 s. We attribute this faster decay rate to the increased photon scattering rate in the higher power sheet potential. We also note here that multimode fibre lasers, as used in this work, have been associated with a power dependent reduction in trap lifetimes, through optical pumping to F=2 states \cite{Will}. However, we do not observe significant deleterious effects at the low powers typically employed in the scanned trap.\\ 


\section{A ring waveguide for BEC}

\begin{figure}[h]
\begin{center}
\includegraphics[width=1\textwidth]{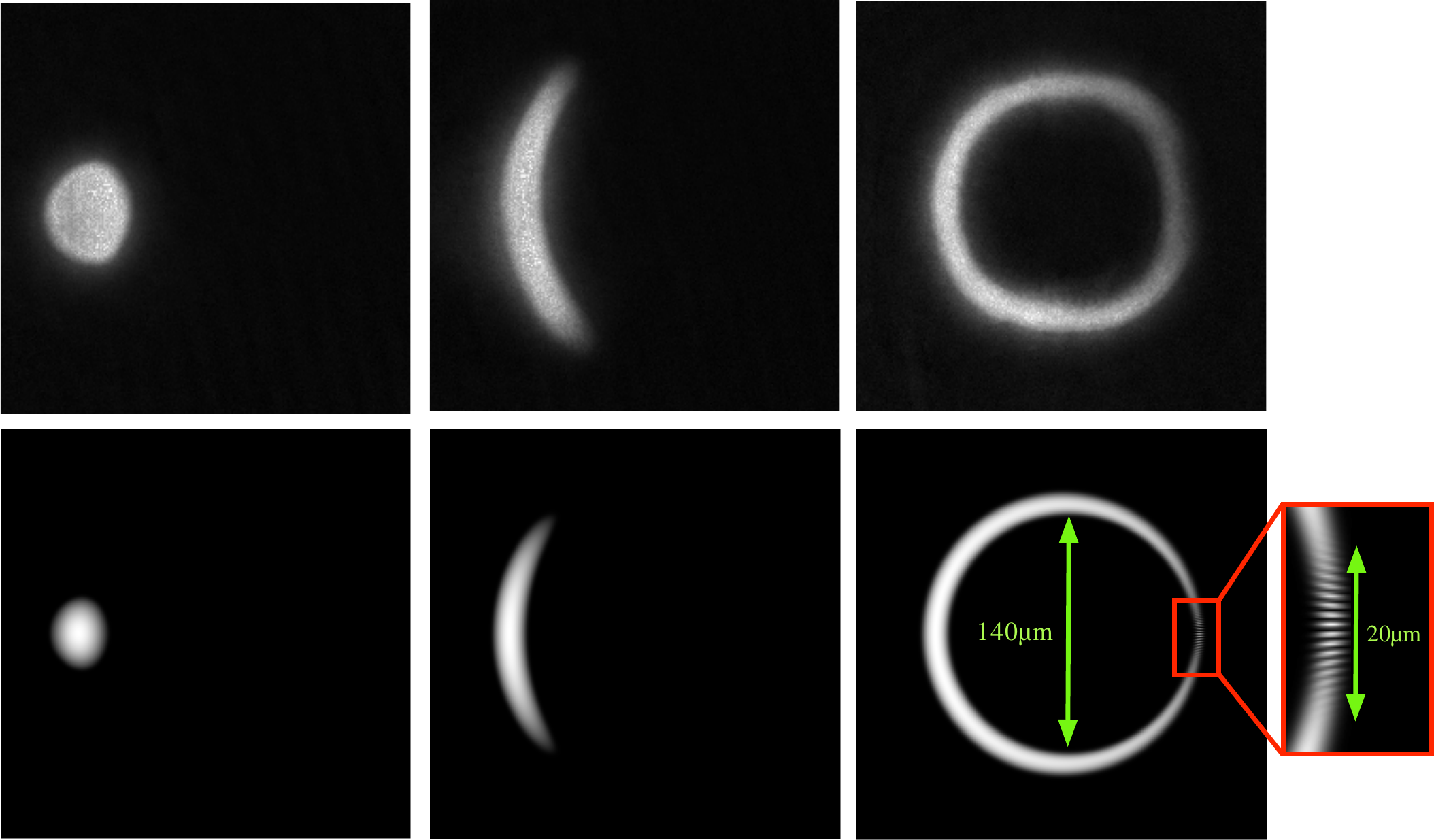}
\end{center}
\caption{{ Expansion and collision of the condensate into a ring waveguide. The upper panels are experimental data at (a) 2.5 ms (b) 12.5 ms (c) 37.5 ms. The lower panels are GPE simulations of $N=10^5$ atoms expanding into the ring potential at the same times. The resulting matter wave interference fringes  (shown at increased magnification to the right) have approximately a 1 $\mu$m spacing. \label{fig5}}}
\end{figure}

As a demonstration of the smoothness of our ring potential we have used the compensated ring trap as a matter waveguide. We initially transfer the BEC from the hybrid trap into the sheet potential where it is pinned using the stationary vertical beam from the AOD. The ring scanning is then triggered, and the mean-field energy of the condensate propels the superfluid atoms around the ring waveguide. We have successfully demonstrated the waveguide with rings up to 300 $\mu$m in diameter \cite{Footnote3}. Figure~\ref{fig5} shows the expansion of the cloud at three different times around the ring, as well as Gross-Pitaevski equation (GPE) simulations \cite{Denn13} of the expanding condensate released from the pin potential into a static ring potential. The dynamics of our condensate match the GPE simulations extremely well, including the characteristic breathing excitation as the condensate expands around the ring. The weak background ring visible in the images is from residual thermal atoms trapped in the sheet potential that are also loaded into the ring. Upon collision on the far side of the ring, the GPE simulation shows the appearance of matter wave fringes, with $\approx$ 1 $\mu$m spacing. Fringes were not visible in the experimental data as they were below the resolution of our imaging system. Future improvements in the imaging system, and reducing the expansion velocity by lowering the mean field energy, should allow the observation of {\emph {in situ}} interference fringes. 

\begin{figure}[h]
\begin{center}
\includegraphics[width=0.8\textwidth]{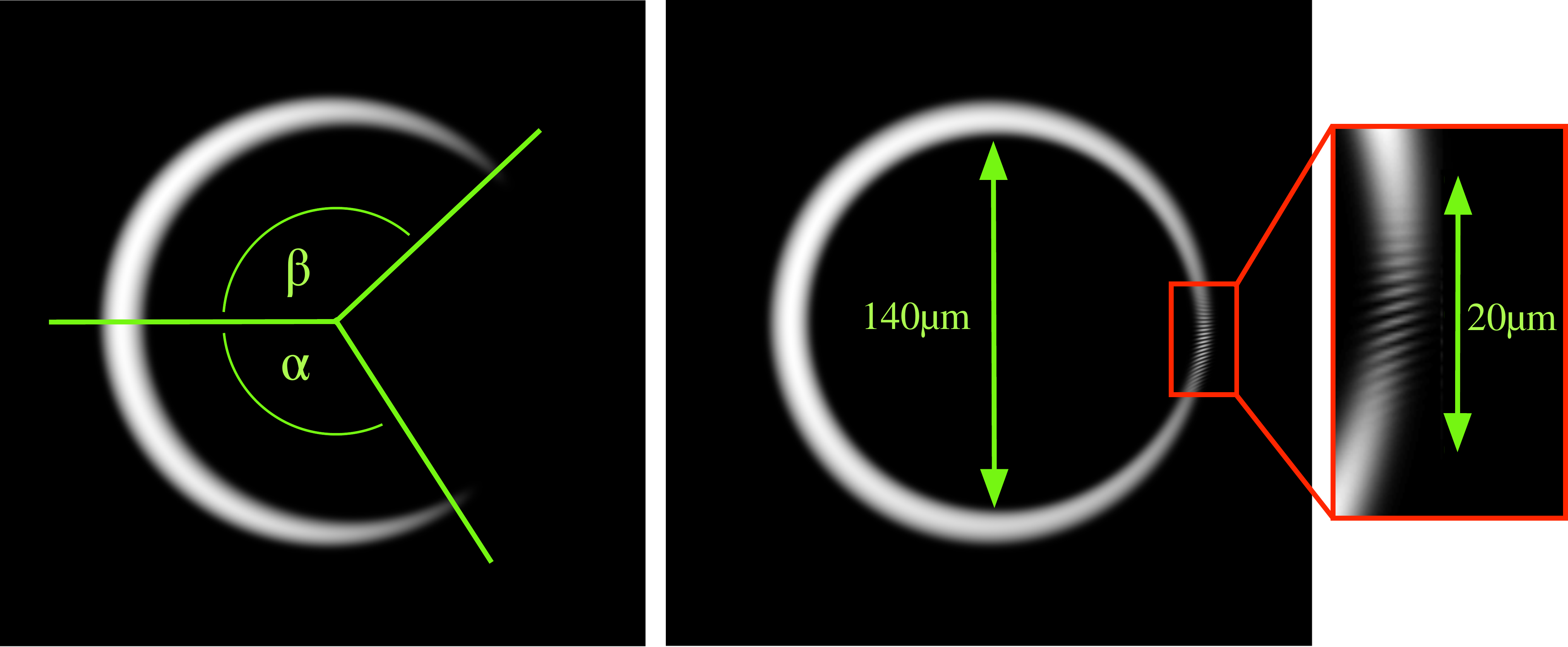}
\end{center}
\caption{{ GPE simulation of the expansion of the condensate into the scanning ring trap (2 kHz). Left: 25 ms of expansion.  Right: 38 ms of expansion.  At 25 ms expansion, the atoms in the upper arm have advanced further (angle $\beta$) around the ring than those in the lower arm  ($\alpha$).  Upon colliding at 38 ms, the two sections of the BEC do not meet head on, and the resulting matter wave interference pattern is distorted and rotated --- see zoomed image on the far right. \label{fig6}}}
\end{figure}

We have also investigated the effect of the time-dependence of the scanning potential on the waveguide, by repeating the GPE simulations and including the switching between the discrete Gaussian potentials (Fig.~\ref{fig6}). At low scan rates (2 kHz) there is a noticeable shift in the cloud expansions in the upper and lower arms, with atoms in the upper arm expanding faster than the lower arm, as can be seen in Fig.~\ref{fig6}(a). The scan in these simulations is in the clockwise direction, with atoms in the upper arm moving with the beam, while in the the lower arm the atoms are moving against the beam scan direction. At higher scan frequencies the difference between the static ring and scanning ring was negligible. This result opens the intriguing possibility of using the scan rate to engineer a precise synthetic rotation on the system, which could be of benefit in benchmarking rotation sensitivity of an interferometer.

\section{Time of flight analysis}

\begin{figure}[h]
\begin{center}
\includegraphics[width=1\textwidth]{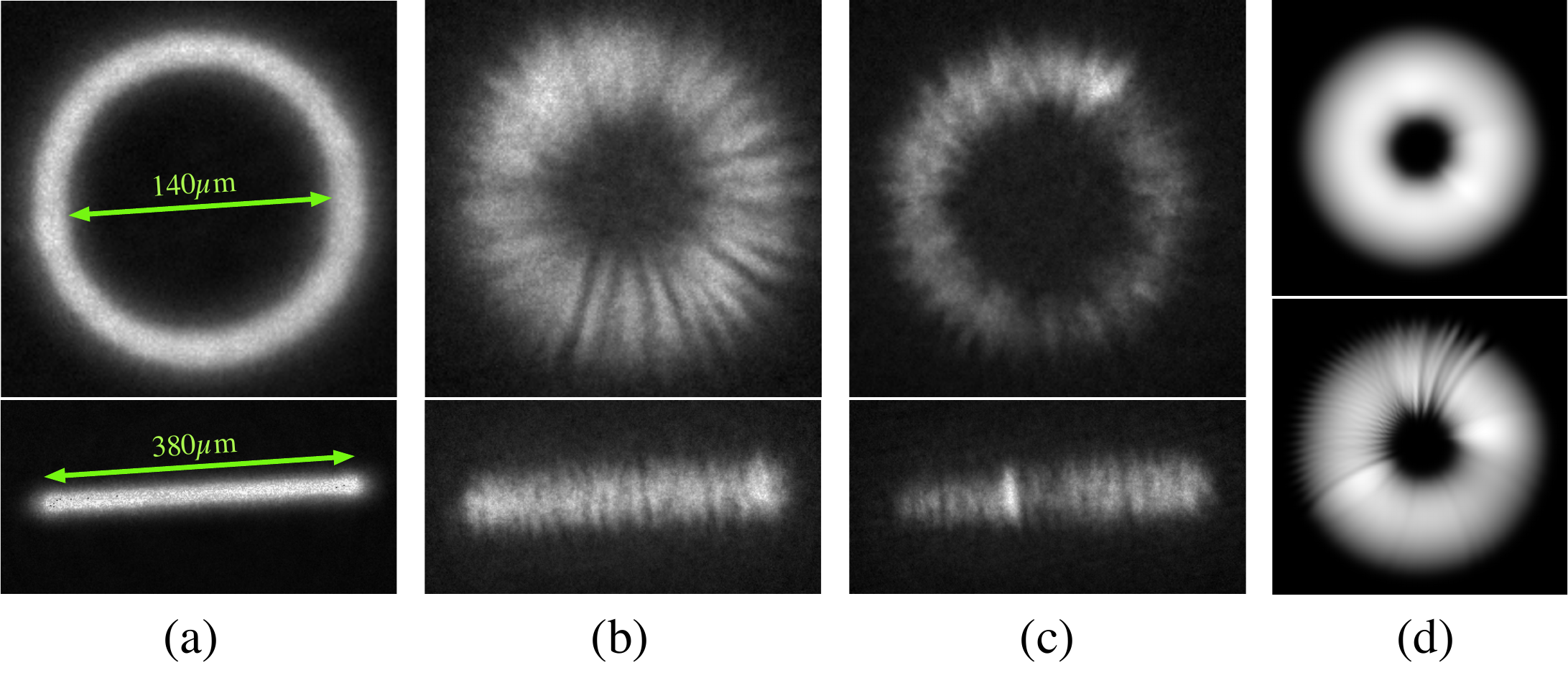}
\end{center}
\caption{{ (a) In-trap images of the BEC in the ring trap (upper) and line trap (lower). (b) Ring trap (upper) and line trap (lower) as for (a) but with 20 ms expansion at 20 kHz scan frequency. (c)  Ring trap (upper) and line trap (lower) as for (a) but with 20 ms expansion at 2 kHz scan frequency.  (d) GPE simulation of atoms in the scanning ring potential with a scan frequency of 8 kHz (upper), and 2 kHz (lower), where excitations are clearly visible.  \label{fig7}}}
\end{figure}

In addition to {\emph{in situ}} imaging of the ring BEC [see Fig.~\ref{fig7}(a)] we have also observed the expansion during time of flight when released from the trap.  A more complex structure in the ring emerges, with fringes in the atom density appearing azimuthally around the ring, and corregations on the inner and outer surface of the ring [see Fig.~\ref{fig7}(b)]. The fringe structure, which we attribute to phonon excitations, is visible even at long hold times of several seconds, showing that the mechanism for their formation  either has very low dissipation or is a continually driven process. Our initial hypothesis was that the excitations were induced by poor mode-matching of the condensate between the hybrid trap and the scanning ring. To test this, we loaded a thermal cloud into the ring potential and allowed it to relax for several seconds, before evaporating the cloud to BEC in the ring by lowering the depth of the sheet potential. However, the fringes were still visible in time of flight. 

The number of fringes appears to be relatively independent of the scan rate of the ring. At high scan rates ($>$ 2 kHz), we additionally changed both the number of points used, and the order in which the pattern was scanned, but did not observe a noticeable change in the pattern. The fringes were observable even at scan rates of 20 kHz. At slower scan rates, the number of fringes appeared to be the same, but there was a clear local density excitation that followed the beam position in the time-averaged potential. This can clearly be seen in Fig.~\ref{fig7}(c) where the ring was scanned at 2 kHz. There is also an observable breathing mode in the ring diameter that can be seen at when scanned at low frequency. The excitations are not peculiar to the ring geometry. We have  loaded the BEC into an intensity compensated line trap of length distance of 380 $\mu$m, and excitations can again be seen in the condensate in time of flight [Fig.~7(a--c)]. Again, at lower scan frequencies (2 kHz) we see localised density of atoms following the position of the scanning beam.

To gain further insight into these excitations we numerically modelled these experiments using the GPE~\cite{Denn13}.  Our initial potential was formed as the sum of twenty-four Gaussian wells with centres spaced evenly around a ring of radius $r_0$ = 50 um, with $x_i$,$y_i$ positions and potential $V(x,y)$ given as
\begin{equation}
{x_i}=-r_0\cos\left(\frac{2\pi}{24}i\right), \qquad{y_i}=r_0\sin\left(\frac{2\pi}{24}i\right),
\end{equation}
\begin{equation}
V(x,y) \propto\sum\limits_{i=0}^{23}\exp\left(-2\frac{[(x-x_i)^2+(y-y_i)^2]}{{w_s}^2}\right).
\end{equation}
After finding the initial ground state in the time-averaged potential, the full time-dependent potential was then simulated for two scan rates, 2 kHz and 8 kHz. Faster scan rates were more computationally expensive due to the smaller time sampling required. The condensate was allowed to evolve in each of these scanning traps for 300 ms, before expanding the condensates in time-of-flight for 20 ms.  At the lower frequency 2 kHz, excitations are readily observable in time of flight, as well as a strong breathing mode [Fig.~7(d), lower panel], consistent with what has been observed experimentally observed. At 8 kHz there are also excitations observable in the simulation, but the amplitude is much reduced [Fig.~7(d), upper panel].  Thus it seems clear that the modulated nature of the time-averaged potential is a contributing factor in driving phonon excitations, particularly at slow scan frequencies.   It seems that in this case the condition $\omega_{scan}\gg\omega_{trap}$ is not  necessarily sufficient to describe the trap as ``time-averaged.''


However, the structures we observe do not seem to be entirely explained by the scanning of the trapping laser, for example see Fig.~\ref{fig7}(b).  We note earlier theoretical work in Modugno \emph{et al.}~\cite{Modu06}, where a BEC in a toroidal trap was simulated while undergoing periodic modulation of the transverse confinement at a frequency $\Omega$=0.6$\omega_{\rho}$. They found this resulted in a longitudinal pattern formation in density around the ring, as a consequence of the amplification of counter-rotating Bogoliubov phonons, with wave vector $k=2 \pi n/{2\pi R}$ where $n$ is the azimuthal angular momentum of the excitation. This obeys the wave equation $c_{s}=\omega(k)/k$, with $c_s$ the speed of sound in the condensate. Experimentally, analogous Faraday wave behaviour was previously observed \cite{Enge07} in extended cigar shaped condensates undergoing transverse modulation near a parametric resonance, resulting in longitudinal pattern formation along the length of the condensate.

Relating our experiments to the work of \cite{Modu06}, if we equate the number $\approx$ 24 of resolvable fringes \cite{Footnote2} in the time-of-flight image to the $2n$ nodes from this model, and using the speed of sound for our condensate $c_s \approx 1$  mm s$^{-1}$, we obtain $\omega (k)\approx \omega_{\rho} = 2\pi \times 47$ Hz. The excitations potentially arise due to a periodic modulation of the radial trap frequency. We note the proximity of $\omega_{\rho}$ to potential external noise sources at \mbox{50 Hz}.  It is thus possible that driving from noise on the trapping light and background AC magnetic fields lead to the excitations.


Finally, we mention the role of phase fluctuations in our system.  It has been known for some time that three-dimensional harmonically trapped Bose gases below the BEC critical temperature may have a phase-coherence length smaller than the 
system size, and hence be in the quasi-condensate regime \cite{Petr01, Garr13}.  These phase fluctuations lead to density stripes in expansion, and have been observed experimentally \cite{Dett01,Shvar02,Rich03}. Recently, Mathey et al \cite{Math10} have considered the quasi-condensate regime for 3D BECs in connected ring trap geometries and found that, as in the case of cigar traps, the requirement for strong phase fluctuations can be expressed in terms of the total number of atoms and the length scales of the trap only. The phase coherence length $l_{\phi}$ is given by 

\begin{equation}
l_{\phi}=\frac{{\hbar}^2 N_0}{k_b T {\pi} m R}
\end{equation}


For our large rings with $R=70 {\mu}$m, and $N_0=1.5 \times 10^5$, at $T=60$ nK, $l_{\phi}=59{\mu}$m ${\approx} L/4$, where $L={\pi}R$ is the characteristic system size (half the ring circumference). The temperature $T_{\phi}$ above which phase fluctuations sare significant is 17 nK, and here $T_{\phi}<T<T_c$, and so our data has been taken in the quasi-condensate regime. The high number of fringes observed in this work is not fully consistent with a cause solely from phase fluctuations, but we cannot fully discount their significance. The phase coherence properties of trapped condensates in time averaged ring potentials is a subject worthy of future study.

\section{Conclusions}

We have demonstrated Bose-Einstein condensates in smooth, intensity compensated ring potentials, and demonstrated long condensate lifetimes in excess of $20$~s. The ring traps are suitable for use as waveguides for a dispersion type interferometer \cite{Kand13}. In time of flight the ring structure exhibits more complex structures, and show excitations that seem in part to be driven by the scanning process.  Another potential cause of these excitations is near-resonant driving of the radial trap frequency by \mbox{50 Hz} noise sources. However, these excitations do not appear to be detrimental to the lifetime of the condensate even at low scan frequencies. 

\section{Acknowledgements}

We would like to thank Prof. Norman Heckenberg, Dr. Stuart Szigeti, Mr Martin Kandes and Prof. Ricardo Carretero-Gonzalez
for useful discussions. This work was supported by the Centre for Engineered Quantum Systems (Grant No. CE110001013), and the Australian Research Council (ARC) Discovery Project Program (Grant No. DP0985142). M. J. D. acknowledges the
support of the JILA Visiting Fellows program and the ARC Discovery Project Program (Grant No. DP1094025). S. A. Haine acknowledges the support of the ARC Discovery Program (Grant No. DE130100575). M. W. J. B. acknowledges the support of an ARC Future Fellowship (FT100100905).


\begin{thebibliography}{10}
\bibitem{Footnote2} For the 140 $\mu$m trap the number of fringes was difficult to resolve exactly in individual images, but appeared to range  between 20--30. By applying a density-density autocorrelation function to the images, and counting the resulting maxima, a mean value of 24 fringes was determined.
\end{thebibliography}
\end{document}